# A glimpse of the end of the dark ages: the $\gamma$-ray burst of 23 April 2009 at redshift 8.3

N. R. Tanvir[1], D. B. Fox[2], A. J. Levan[3], E. Berger[4], K. Wiersema[1], J. P. U. Fynbo[5], A. Cucchiara[2], T. Krühler[6,7], N. Gehrels[8], J. S. Bloom[9], J. Greiner[6], P. Evans[1], E. Rol[10], F. Olivares[6], J. Hjorth[5], P. Jakobsson[11], J. Farihi[1], R. Willingale[1], R. L. C. Starling[1], S. B. Cenko[9], D. Perley[9], J. R. Maund[5], J. Duke[1], R. A. M. J. Wijers[10], A. J. Adamson[12], A. Allan[13], M. N. Bremer[14], D. N. Burrows[2], A. J. Castro Tirado[15], B. Cavanagh[12], A. de Ugarte Postigo[16], M. A. Dopita[17], T. A. Fatkhullin[18], A. S. Fruchter[19], R. J. Foley[4,20], J. Gorosabel[15], S. T. Holland[8], J. Kennea[2], T. Kerr[12], S. Klose[21], H. A. Krimm[22,23], V. N. Komarova[18], S. R. Kulkarni[24], A. S. Moskvitin[18], T. Naylor[13], B. E. Penprase[25], M. Perri[26], P. Podsiadlowski[27], K. Roth[28], R. E. Rutledge[29], T. Sakamoto[22], P. Schady[30], B. P. Schmidt[17], A. M. Soderberg[4], J. Sollerman[5,31], A. W. Stephens[28], G. Stratta[26], T. N. Ukwatta[8,32], D. Watson[5], E. Westra[4], T. Wold[12], C. Wolf[27]

[1]*Department of Physics and Astronomy, University of Leicester, University Road, Leicester, LE1 7RH, UK*

[2]*Department of Astronomy & Astrophysics, Pennsylvania State University, University Park, PA 16802, USA*

[3]*Department of Physics, University of Warwick, Coventry, CV4 7AL, UK*

[4]*Harvard-Smithsonian Center for Astrophysics, 60 Garden Street, Cambridge, MA 02138, USA*

[5]*Dark Cosmology Centre, Niels Bohr Institute, University of Copenhagen, Juliane Maries Vej 30, 2100 Copenhagen, Denmark*

[6]*Max-Planck-Institut für Extraterrestrische Physik, Giessenbachstraße 1, 85740 Garching, Germany*

[7]*Universe Cluster, Technische Universitat München, Boltzmannstrasse 2, 85748 Garching, Germany*

[8]*NASA Goddard Space Flight Center, Greenbelt, MD 20771, USA*

[9]*Department of Astronomy, University of California, Berkeley, CA 94720-3411, USA*

[10]*Astronomical Institute "Anton Pannekoek", University of Amsterdam, PO Box 94249, 1090 GE Amsterdam , The Netherlands*

[11]*Centre for Astrophysics and Cosmology, Science Institute, University of Iceland, Dunhagi 5, 107 Reykjavík, Iceland*

[12]*Joint Astronomy Centre, 660 N. A'ohoku Place, University Park, Hilo, Hawaii 96720, USA*

[13]*School of Physics, University of Exeter, Stocker Road, Exeter, EX4 4QL, UK*

[14]*H H Wills Physics Laboratory, University of Bristol, Tyndall Avenue, Bristol BS8 1TL, UK*

[15]*Instituto de Astrofísica de Andalucía del Consejo Superior de Investigaciones Científicas (IAA-CSIC), PO Box 03004, E-18080 Granada, Spain*

[16]*European Southern Observatory, Casilla 19001, Santiago, Chile*

[17]*Research School of Astronomy & Astrophysics, The Australian National University, Cotter Road, Weston Creek ACT 2611, Australia*

[18]*Special Astrophysical Observatory, Nizhnij Arkhyz, Karachai-Cirkassian Republic,* 369167 Russia

[19]*Space Telescope Science Institute, 3700 San Martin Drive, Baltimore, MD21218, USA*

[20]*Clay Fellow*

[21]*Thüringer Landessternwarte Tautenburg, Sternwarte 5, D-07778 Tautenburg, Germany*

[22]*CRESST and NASA Goddard Space Flight Center, Greenbelt, MD 20771, USA*

[23]*Universities Space Research Association, 10211 Wincopin Circle, Suite 500, Columbia, MD 21044, USA.*

[24]*Department of Astronomy, California Institute of Technology, MC 249-17, Pasadena, CA 91125, USA*

[25]*Department of Physics and Astronomy, Pomona College, Claremont, CA 91711, USA*

[26]*ASI Science Data Center, via Galileo Galilei, 00044 Frascati, Italy*

[27]*Department of Physics, Oxford University, Keble Road, Oxford, OX1 3RH, UK*

[28]*Gemini Observatory, Hilo, HI 96720, USA*

[29]*Physics Department, McGill University, 3600 rue University, Montreal, QC H3A 2T8, Canada*

[30]*The UCL Mullard Space Science Laboratory, Holmbury St Mart, Dorking, Surrey, RH5 6NT, UK*

[31]*The Oskar Klein Centre, Department of Astronomy, Stockholm University, 106 91 Stockholm, Sweden*

[32]*The George Washington University, Washington, DC. 20052, USA*

**It is thought that the first generations of massive stars in the Universe were an important, and quite possibly dominant[1], source of the ultra-violet radiation that reionized the hydrogen gas in the intergalactic medium (IGM); a state in which it has remained to the present day. Measurements of cosmic microwave background anisotropies suggest that this phase-change largely took place[2] in the redshift range $z$=10.8 ±1.4, while observations of quasars and Lyman-α galaxies have shown that the process was essentially completed[3,4,5] by $z$≈6. However, the detailed history of reionization, and characteristics of the stars and proto-galaxies that drove it, remain unknown. Further progress in understanding requires direct observations of the sources of ultra-violet radiation in the era of reionization, and mapping the evolution of the neutral hydrogen (H I) fraction through time. The detection of galaxies at such redshifts is highly challenging, due to their intrinsic faintness and high luminosity distance, whilst bright quasars appear to be rare beyond $z$≈7 [ref 6]. Here we report the discovery of a gamma-ray burst, GRB 090423, at redshift $z = 8.26^{+0.07}_{-0.08}$. This is well beyond the redshift of the most distant spectroscopically–confirmed galaxy ($z$=6.96; ref 7) and quasar ($z$=6.43; ref 8). It establishes that massive stars were being produced, and dying as GRBs, ~625 million years after the Big Bang. In addition, the accurate position of the burst pinpoints the location of the most distant galaxy known to date. Larger samples of GRBs beyond $z$~7 will constrain the evolving rate of star formation in the early universe, while rapid spectroscopy of their afterglows will allow direct exploration of the progress of reionization with cosmic time[9,10].**

It has long been recognised that GRBs have the potential to be powerful probes of the early universe. Known to be the end product of rare massive stars[11], GRBs and their afterglows can briefly outshine any other source in the universe, and would be theoretically detectable to $z \sim 20$ and beyond[12,13]. Their association with individual stars means that they serve as a signpost of star formation, even if their host galaxies are too faint to detect directly. Equally important, precise determination of the hydrogen Lyman-$\alpha$ absorption profile can provide a measure of the neutral fraction of the IGM at the location of the burst[9,10,14,15]. With multiple GRBs at $z > 7$, and hence lines of sight through the IGM, we could thus trace the process of reionization from its early stages. However, until now the highest redshift GRBs (at $z = 6.3$ [ref 16] and 6.7 [ref 17]) have not exceeded other cosmological sources.

GRB 090423 was detected by the Burst Alert Telescope (BAT) on the *Swift* satellite[18] at 07:55:19 UT on 23 April 2009. Observations with the on-board X-ray Telescope (XRT), beginning 73 s after burst trigger, revealed a variable X-ray counterpart and localized its position to a precision of 2.3 arcsec (90% confidence). Ground-based optical observations in the *r, i* and *z* filters starting within a few minutes of the burst revealed no counterpart at these wavelengths (see Supplementary Information (SI)).

The United Kingdom Infrared Telescope (UKIRT) in Hawaii responded to an automated request, and began observations in the *K*-band 21 minutes post burst. These images (Figure 1) revealed a point source at the reported X-ray position, which we concluded was likely to be the afterglow of the GRB. We also initiated further near-infrared (NIR) observations using the Gemini-North 8-m telescope, which started 75 min after the burst, and showed that the counterpart was only visible in filters redder than about 1.2 μm. In this range the afterglow was relatively bright and exhibited a shallow spectral slope $F_\nu \propto \nu^{-0.26}$, in contrast to the deep limit on any flux in the *Y* filter (0.97–1.07 μm). Later observations from Chile using the MPI/ESO 2.2m telescope, Gemini South and the Very Large Telescope (VLT) confirmed this finding. The non-detection in the *Y*-band implies a power-law spectral slope between *Y* and *J* steeper than $F_\nu \propto \nu^{-18}$. This is impossible for dust at any redshift, and is a text-book case of a short-wavelength “drop-out” source. It shows that a sharp spectral break occurs around 1.1–1.2 μm; the full *grizYJHK* SED obtained ~17 hours after burst gives a photometric redshift $z = 8.06^{+0.21}_{-0.28}$, assuming a simple IGM absorption model**.** Complete details of our imaging campaign are given in SI Table 1.

Our first NIR spectroscopic observations were made with the European Southern Observatory (ESO) Very Large Telescope (VLT-UT1) 8.2-m telescope in Chile, starting about 17.5 hours post-burst. These covered 0.98–1.1 μm (45 min exposure) and 1.1–1.4 μm (35 min exposure) and revealed no emission blue-ward of a step at about 1.14 μm, confirming the origin of the break as being due to Lyman-α absorption by neutral hydrogen, with a redshift of $z$~8.3. The spectrum and broadband photometric observations, over-plotted with a damping wing model, are shown in Figure 2. To obtain a more quantitative estimate of the redshift, we must make some assumption about the neutral hydrogen column density ($N_{HI}$) of the GRB host galaxy. A larger $N_{HI}$ will mean a broad damping wing and a lower redshift, as shown by the confidence contours in the inset panel of Figure 2. For simplicity we take a flat prior likelihood between 19 and 23 for the $\log(N_{HI}/cm^{-2})$ of the host, which is broadly consistent with the distribution observed for GRB hosts[19,20], and further assume the neutral fraction of the IGM to be 10% (although our conclusions depend only weakly on this). Hence we find the redshift from the ISAAC spectroscopy to be $z = 8.20^{+0.04}_{-0.07}$. An additional spectrum, taken ~40 hours post burst with VLT/SINFONI (2.5 hour exposure) independently confirms this analysis, yielding $z = 8.33^{+0.06}_{-0.11}$. Fitting simultaneously to both spectra and photometric data points gives our best estimate of the redshift $z = 8.26^{+0.07}_{-0.08}$. The low signal-to-noise ratio means we are unable to detect metal absorption features in either spectrum, which would otherwise provide a more precise value of the redshift, and also prevents a meaningful attempt to measure the IGM H I column density in this instance. Our three independent redshift measures are broadly consistent with that reported from a low-resolution spectrum obtained with the Telescopio Nazionale Galileo (TNG) on La Palma[21].

The X-ray and NIR light curve of GRB 090423 is shown in Figure 3. The X-ray exhibits a large flare peaking at about 200 s, which precedes a plateau phase. The plateau, which is also seen in the infrared, breaks into a power-law decline at about 1.5 hours, which corresponds to about 10 minutes in the rest frame. The spectral slope seen in the NIR is considerably shallower than that seen in the X-ray ($F_{\nu,\mathrm{xray}} \propto \nu^{-1}$) consistent with a cooling break between the two bands (see SI for further description). Apart from the unusually blue continuum between *J* and *K*, this behaviour is otherwise unexceptional for GRBs observed at lower redshifts.

With the standard cosmological parameters ($H_0 = 71$ km s$^{-1}$ Mpc$^{-1}$; $\Omega_M = 0.27$; $\Omega_\Lambda = 0.73$) a redshift of $z = 8.26$ corresponds to a time of only 625 million years after the Big Bang, when the Universe was just 4.6% of its current age. GRB090423's observed fluence, coupled with its luminosity distance of 85.4 Gpc, indicates this event was a bright, but not extreme, GRB, with an implied isotropic equivalent energy of $E_{\mathrm{iso}} = 1 \times 10^{53}$ erg (8–1000 keV)[22]**.** This, and other observed properties of GRB 090423 (such as its afterglow characteristics and peak energy) are generally consistent with the bulk of the GRB population. Thus we find no evidence of exceptional behaviour which might indicate an origin from a population III progenitor. It is expected that first generation stars are more likely to collapse to particularly massive black-holes that in turn may produce unusually long-lived GRBs[23], which is certainly not the case for GRB 090423.

Indeed, it is noteworthy that the γ-ray duration of GRB 090423 of $t_{90} = 10.3$ seconds, corresponds in the rest-frame to only 1.1 seconds. Two other $z > 5$ GRBs (060927 and 080913) had similarly short rest-frame durations, leading to some debate[24] as to whether their progenitors are similar to those of short-hard GRBs, which are not thought to be directly related to core-collapse. However, in the case of GRB 090423, a more careful extrapolation of the observed γ- and X-ray lightcurve to lower redshift shows that its duration would have appeared significantly longer than suggested by naïve time-dilation considerations[25]. In any event, the progenitors of short GRBs likely involve compact objects that are themselves the end products of massive stars, so the above conclusions will hold irrespective of the population from which GRB 090423 derives.

Beyond breaking a distance record, the high redshift of GRB 090423 has several crucial implications. First, it demonstrates that massive stars did indeed exist at that time. Predictions based on extrapolating the global star-formation rate to higher redshifts, suggest that the rate of GRBs at $z\approx8$ should be about 40% of that at $z\approx6$ [ref 23] and that approximately 2% of GRBs above the Swift detection threshold should come from $z>8$. Given the extra difficulty of identifying afterglows at higher redshifts, our finding is broadly consistent with these predictions. This is extremely encouraging for the prospects of future space missions optimised for finding high-redshift GRBs, and ultimately for using them to measure the history of star formation at very high redshifts[26]. Second, it is close to the redshift range during which the bulk of the cosmic reionization is thought to have taken place. GRBs at a similarly high redshift for which infrared spectroscopy was possible earlier, or which had brighter afterglows, would provide a direct probe of the progress of reionization[10]. This is not an unreasonable hope: the most extreme GRBs have had afterglows that were intrinsically significantly brighter at the same rest-frame time[13,27], while our first spectra were only secured more than 15 hours after the burst. High signal-to-noise spectroscopy would provide a measure of the hydrogen neutral fraction in the IGM in the vicinity of the burst, and also the metallicity of the host galaxy, which potentially offers important clues to the nature of any earlier generations of stars. Since the massive stars which yield GRBs are also likely to belong to the same population that is responsible for reionization, this suggests that GRBs will ultimately be used to constrain both the supply and demand side of the cosmic ionization budget at early times.

Future deep observation of the host galaxy of GRB 090423, and comparison of its properties to those of known $z \approx 6$ systems[28,29] may provide stronger constraints on the nature of the underlying stellar population, and hence on the possibility that GRB 090423 originated in a galaxy undergoing its first burst of star formation. If the host is relatively bright compared to those of other GRBs[28,30] it may be detectable with the refurbished Hubble Space Telescope + WF3 ($H_{AB}$≈28 mag); detection of a more typical GRB host galaxy ($H_{AB}$≥29 mag) will have to await the advent of the James Webb Space Telescope.

'**Supplementary Information** accompanies the paper on **www.nature.com/nature**.'

The authors confirm they have no competing financial interests.

Correspondence and requests for materials should be addressed to N.R.T. (e-mail: nrt3@star.le.ac.uk).

Acknowledgements:

We thank Ph. Yock, B. Allen, P. Kubanek, M Jelinek and S. Guziy for their assistance with the BOOTES-3 YA telescope observations. Based on observations obtained at the Gemini Observatory, which is operated by the Association of Universities for Research in Astronomy, Inc., under a cooperative agreement with the NSF on behalf of the Gemini partnership: the National Science Foundation (United States), the Science and Technology Facilities Council (United Kingdom), the National Research Council (Canada), CONICYT (Chile), the Australian Research Council (Australia), Ministério da Ciência e Tecnologia (Brazil) and SECYT (Argentina). Based on observations made with ESO Telescopes at the La Silla or Paranal Observatories under programme ID 083.A-0552. The UKIRT is operated by the Joint Astronomy Centre on behalf of the UK Science and Technology Facilities Council.

All authors made contributions to this paper. This took the form of triggering observations (NRT, DBF, AJL, EB, JSB, DP, JG, AJCT, ADUP); direct analysis of ground based data (NRT, DBF, AJL, EB, KW,

JPUF, AC, JSB, JF, JD, JG, BC, DP, JM, KM, TK, AJCT, AD); Swift analysis (PAE, RLCS, AJL, NRT, NG, DW, PS, TS); observations at various observatories and their automation to accept GRB overrides (AJA, AA, TK, TN, AS, KR, TW). Additionally, all authors have made contributions through their major involvement in the programmes from which the data derives, and in contributions to the interpretation, content and discussion presented here.

**Figure 1.** Multiband images of the afterglow of GRB 090423. The right panel was the discovery image made with the UKIRT Wide Field Camera (WFCAM) in K at a mid-time of about 30 minutes post-burst. The other three images were obtained approximately 1.5 hours post-burst with the Gemini-N Near Infrared Imager and Spectrometer (NIRI). The main panels are approximately 40 arcsec on a side, with north up and east left. In each case the inset shows the region around the GRB, smoothed and at high contrast. The non-detection in the Y-band (1.02 μm) to deep limits, combined with clear detection in J (1.26 μm), H (1.65 μm) and K (2.15 μm) gave the first strong indication of a very high redshift source. The absence of any flux in Y, coupled with the blue colour in the longer wavelength bands (J-H(AB)~0.15 mag) cannot be explained by any dust extinction model, and implies a redshift greater than about 7.8 for GRB 090423.

**Figure 2.** The SZ- and J-band 1D and 2D spectra of GRB 090423 obtained with the VLT using the Infrared Spectrometer And Array Camera (ISAAC). Also plotted are the sky-subtracted photometric data points obtained with the Gemini-N/NIRI (red) in Hawaii, and VLT/HAWKI and Gemini-S/GMOS (blue) in Chile (in all cases scaled to 16 hours post burst). The vertical error bars are 2σ (i.e. 95% confidence), whilst the horizontal lines indicate the widths of the filters. The shorter wavelength measurements are, of course, non-detections, and emphasise the tight constraints on any transmitted flux below the break. The break itself, at about 1.14 μm, is seen to occur close to the short-wavelength limit of the J-band spectrum, below which, although noisy, there is no evidence of any detected continuum. Details of the data reduction steps and adaptive binning used to construct these spectra are given in the Supplementary

Information. A model spectrum showing the H I damping wing for a host galaxy with hydrogen column density $N_{HI}=10^{21}$ cm$^{-2}$ at a redshift $z$=8.26 is overplotted (solid black line), and provides a good fit to the data. Allowing for a wider range in possible host $N_{HI}$ gives the 1$\sigma$ (68%) and 2$\sigma$ (95%) confidence contours shown in the inset panel. Interestingly, the fact that no deviation is seen from a power-law spectrum redward of 1.2 $\mu$m together with its blue slope suggests that there is little or no dust along the line of sight through the GRB host galaxy (unless it is “grey”), consistent with the galaxy being relatively un-evolved, and with a low abundance of metals.

**Figure 3.** The X-ray (top) and infrared (bottom) lightcurves of GRB 090423 showing both observed (left and bottom axes) and rest-frame (right and top axes) properties. The X-ray lightcurve was obtained by the Swift Burst Alert Telescope (BAT – cyan) and X-ray Telescope (XRT -magenta), where the BAT observations have been extrapolated into the X-ray band assuming the observed spectral model. The overall fit consists of a prompt plus afterglow model, fit simultaneously to both BAT and XRT observations. The IR lightcurve was obtained by UKIRT, Gemini-North the MPI/ESO 2.2m telescope and the VLT. For consistency, although individual bands are plotted they have been transformed into absolute magnitudes in the J-band based on our best fitting SED, which has slope $F_\nu \propto \nu^{-0.26}$. Although at first sight similar, the two lightcurves differ somewhat in their detail. In particular the break time between shallow and steeper decline appears to occur later in the optical/IR, although the post-break slopes are similar ($F_t \propto t^{-1.35}$). Further the decline in the NIR shows more variation around the power-law decline, being significantly in excess at the 4$^{th}$ epoch of observation, around 300 000 s, possibly indicative of a flare. Note that the absolute magnitude scale corresponds to 0.136 μm AB-magnitudes. For further details please see the Supplementary Information.

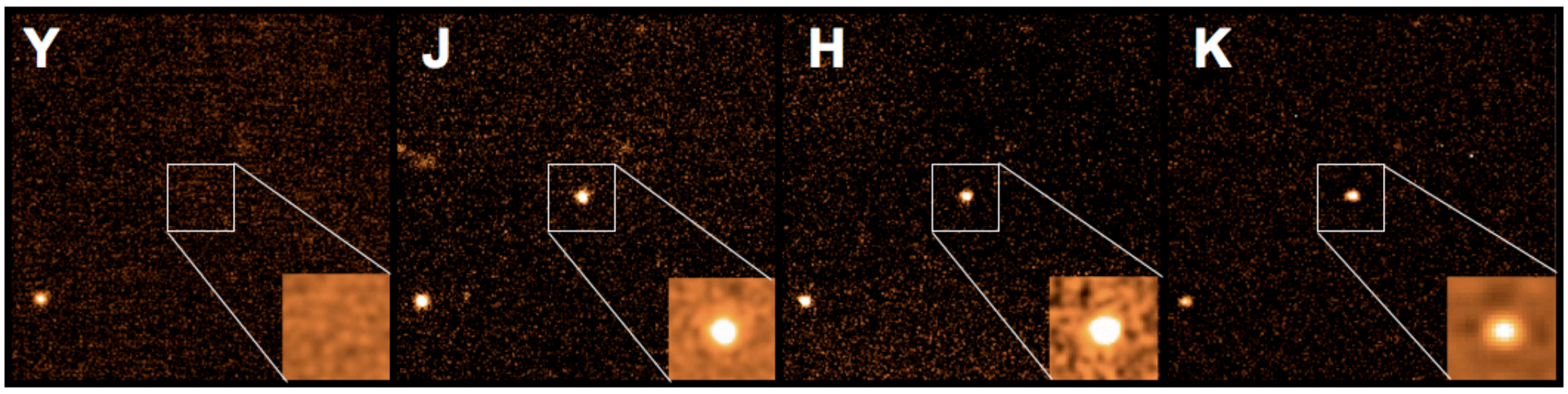

**Figure 1.**

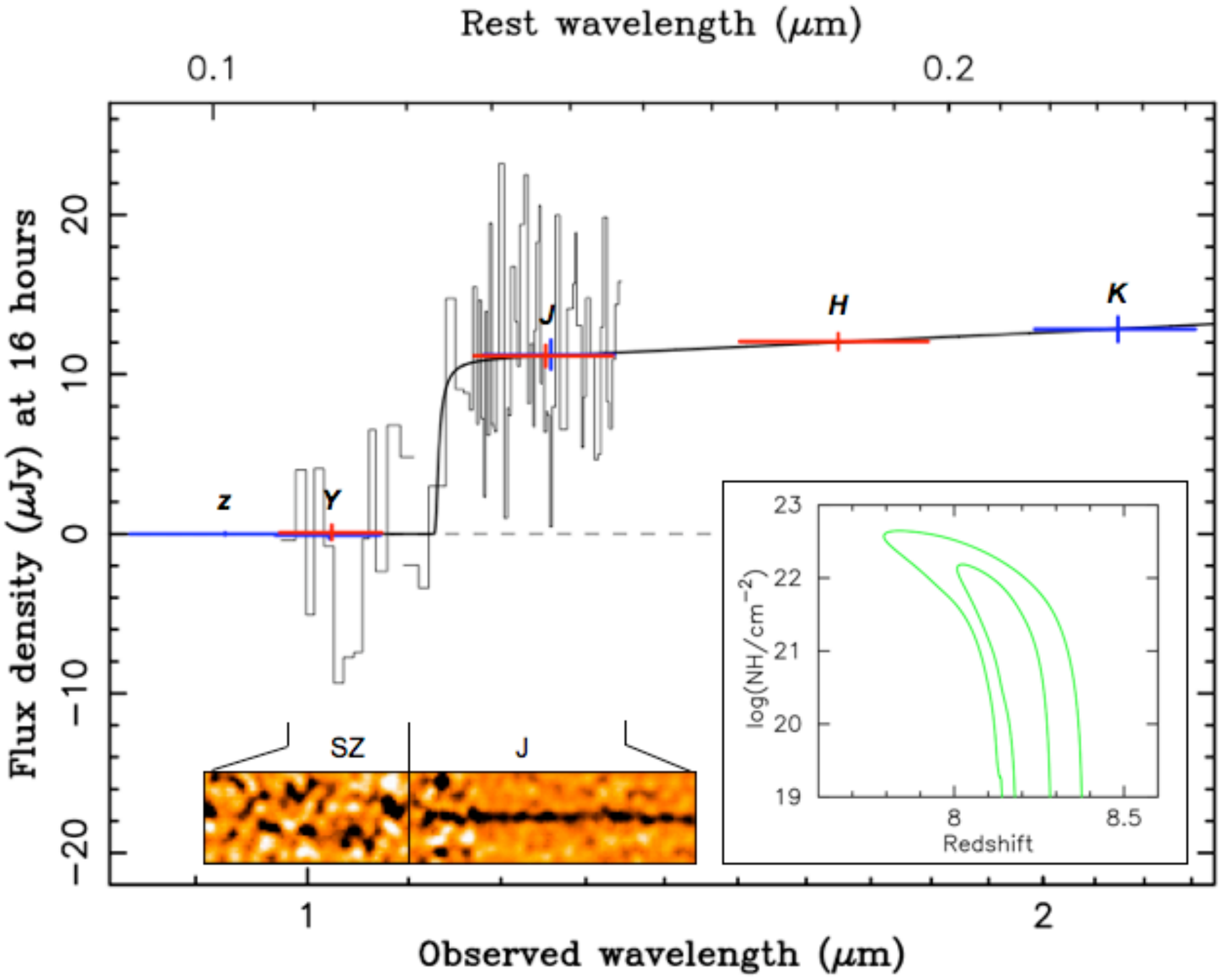

**Figure 2.**

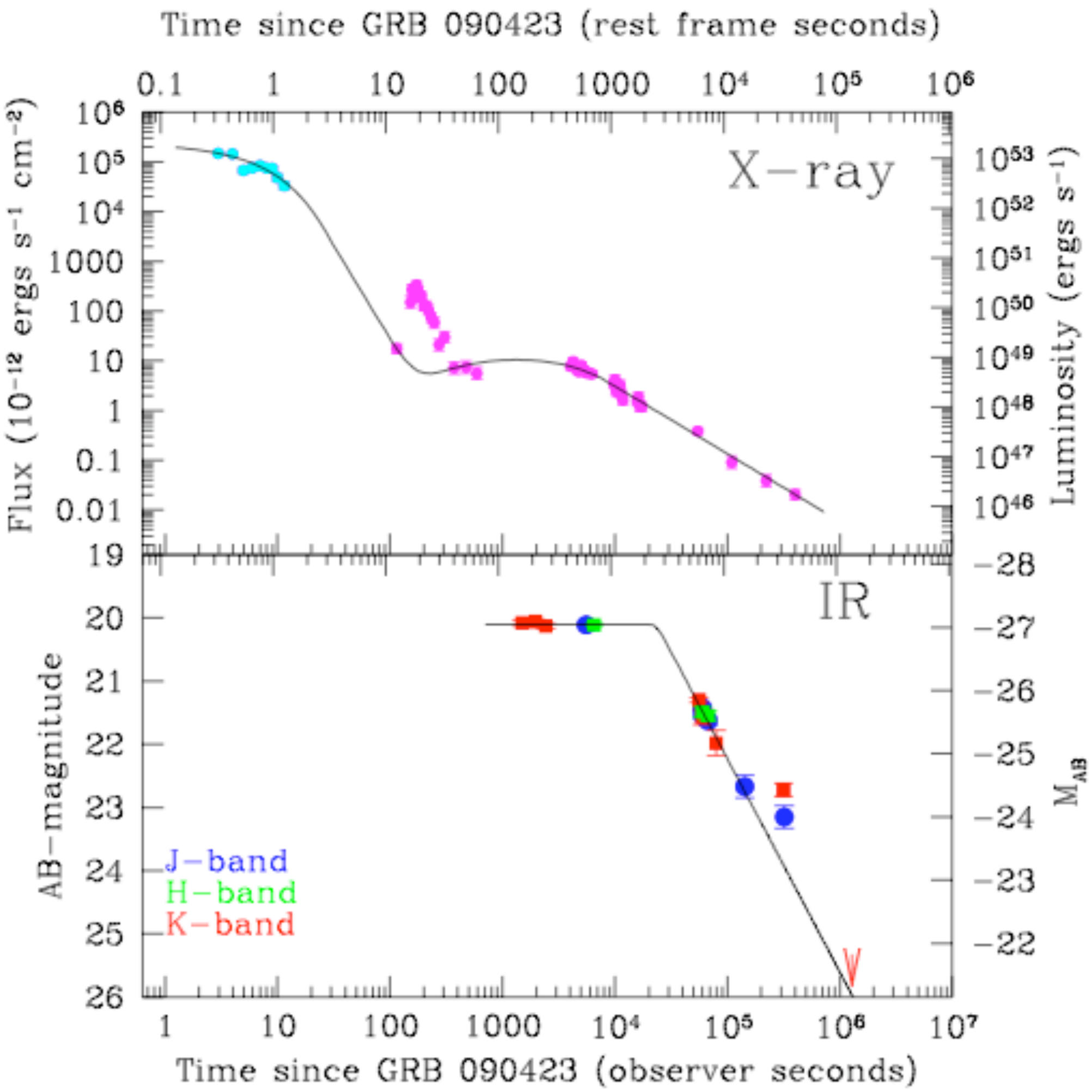

**Figure 3.**

# Supplementary Information:

# A glimpse of the end of the dark ages: the γ-ray burst of 23 April 2009 at redshift 8.3

## GROUND BASED OBSERVATIONS

### *Optical and Infrared Imaging*

Our campaign of optical and IR observations of GRB 090423 began roughly 40 seconds post burst, and continued for ~20 days. The facilities used in this campaign included with the Burst Observer and Optical Transient Observing System (BOOTES-3) Yock-Allen telescope in New Zealand, the Palomar 60 inch[31], the 3.9m United Kingdom Infrared Telescope (UKIRT, equipped with the eStar system[32]), the Gemini North and South Observatories, the Russian 6.0m BTA, the Max-Planck-Institute/ESO 2.2m Telescope and GRB Optical and NIR Detector (GROND) 7-channel imager[33] and the European Southern Observatories (ESO) Very Large Telescope (VLT), Antu and Yepun (UT1 and UT4) units. Imaging observations have been reduced using the standard techniques. Aperture photometry has been performed, and results are given relative to the nearby bright star at RA=09:55:35.31, Dec =18:09:30.9, which has tabulated magnitudes of $r$=17.778(0.006), $i$=16.537(0.005), $z$=15.905(0.004), $Y$=15.25(0.02), $J$=14.523(0.031), $H$=13.972(0.036), $K$=13.773(0.049), (numbers in parenthesis correspond to 1-sigma errors). Optical magnitudes for this star are derived from Sloan Digital Sky Survey (SDSS) observations[34], while nIR measurements are obtained by the 2-Micron All Sky Survey (2MASS)[35]. Our $Y$-band measurements come from the field calibration of Im et al[36]. Colour terms for our different filters are minimal and are not considered. Upper limits are stated at 3-sigma confidence levels. In addition, the fluxes shown correspond to sky subtracted flux, measured in an aperture centred on the location of the afterglow. In observations where the afterglow is not visible its location has been derived by relative astrometry between filters where the object is bright, and those where it is undetected. Roughly 10 point sources common to each image were used for this alignment in the Image Reduction and Analysis Facility (IRAF) task *geomap*. The resulting astrometric fits are better than 0.05 arcseconds (better for cases of aligning images taken with the same instrument). A mosaic, showing multicolour images of the GRB 090423 afterglow taken at three separate epochs is shown in Figure S1, while a full log of all photometric observations considered in this paper is included in Table S1.

*Infrared Spectroscopy*

Near-infrared spectroscopy was first obtained using the VLT Imaging Spectrometer and Array Camera (ISAAC). Data were obtained using the J grism (~1.08 - 1.32 micron; midpoint 17.94 hr after burst) for a total of 36 min exposure time, and in the SZ grism (~0.92 - 1.20 micron) for a total exposure time of 45 min. The data were acquired in a large (60") nodding pattern along the slit, with individual exposures of 60 s, to facilitate sky subtraction, while telluric standard stars were observed at similar airmass, immediately following the science observations. The data were processed using the ESO / ISAAC pipeline, which includes sky subtraction, flat fielding, wavelength calibration, spectral image registration and co-addition. Finally we rectified the spectra to produce horizontal traces, and removed a 50 Hz pickup pattern. The fully reduced science frame of the J-grism data shows a clear detection of the source, whereas no trace is visible in the reduced SZ-grism frame. The spectrum of the afterglow was obtained with the IRAF appall task, using the trace of the telluric standard as a reference aperture. Telluric correction and flux calibration was achieved using the IDL / SpeXtool package *xtellcor_general*[37], using a high quality Vega model convolved with known telluric absorption features to match the entire telluric standard (HD 58784, B3III) spectrum. This model is then applied to the science target spectrum, producing a telluric feature corrected and fully flux calibrated spectrum (excepting slit losses); the data are completely corrected for instrument sensitivity. A small (3 km/s) wavelength shift was applied to the telluric standard to minimize residual noise near significant telluric features. The final flux calibration of the J-grism spectrum was achieved using the J-band photometry of the afterglow.

We obtained further near-infrared spectroscopy on the 24$^{th}$ April, using the VLT Spectrograph for Integral Field Observations in the Near Infrared (SINFONI). Data were obtained using the J grism, providing a spectral resolution of ~2000, using a 8" x 8" field of view. Fifteen exposures, each of 600 seconds were obtained, with midpoint 1.276 days after burst. We reduced the data using the ESO esorex SINFONI pipelines, version 1.9.4. This includes correction for detector signatures: bad pixels, detector contribution to the measured signal, dark current, flat fielding (correct pixel to pixel gain variations and relative slitlet throughput differences) and geometric distortions. Wavelength calibration was performed using arc spectra as well as OH sky lines. After these reductions, the image slices were combined using the spatial and spectral information to form a 3-dimensional datacube. In the co-added cube a source is visible when adding together all wavelength slices. We extracted a 1-dimensional spectrum of this source by using a 5 pixel radius Gaussian-weighted aperture at the source position. To correct for some residual sky emission lines and thermal signatures, we extracted sky spectra from the same cube (besides the afterglow no other sources are visible in the 8"x8" field of view), and subtracted this sky spectrum from the science spectrum. We reduced and extracted the spectrum of the telluric standard star (HD 88322, type B4V) in the same manner, using the same routines to correct for telluric transmission and flux calibrate as for the ISAAC spectroscopy. Though the signal-to-noise is low, the high resolution of the SINFONI spectroscopy offers larger windows between sky lines, and the final spectrum plotted (Figure S05) shows 100 Angstrom wide bins in which the flux in the target aperture has been coadded using variance-weighting to down-weight frames badly affected by high sky background.

Both ISAAC and SINFONI spectroscopy show clear breaks at around 1.14 microns. More detailed spectral fitting, and the constraints that this yields on the redshift are presented below. A log of the spectroscopic observations is shown in Table S2.

## SWIFT OBSERVATIONS

The *Swift* Burst Alert Telescope (BAT) triggered on GRB 090423 at 07:55:19 UT on 23 April 2009, yielding a prompt position with a 3 arcminute error radius (90% confidence)[38]. Observations with the narrow field instruments (the X-ray Telescope (XRT) and UV and Optical Telescope (UVOT)) began approximately 70 seconds after the burst. Observations with the XRT ultimately yielded a position accurate to 1.7 arcseconds[39,40]. The UVOT observations yielded no detection as expected for a source at z~8 (ref 41). A summary of the results of the BAT and XRT observations is provided in Table S3, while more details are given in the next sections.

### Properties of prompt emission

The prompt gamma-ray lightcurve of GRB 090423 consists of two peaks with a total duration $t_{90}$ of 10.3 ± 1.1 seconds, with a fluence of $5.9 \pm 0.4 \times 10^{-7}$ ergs $cm^{-2}$ (15-150 keV). Its prompt spectrum can be fit with either a Band function[42] or a cut-off power-law, which yields a peak energy ($E_p$) of ~50 keV [ref 43,44]). In particular, comparing the properties of the prompt emission of GRB 090423 with those of lower-z GRBs shows no obvious differences that may be expected from time dilation or spectral softening as the harder photons are redshifted. Hence, in terms of these basic properties

GRB 090423 is broadly consistent with the broader GRB population observed by *Swift*, and does not have obvious features that would allow it to be identified as a high-*z* candidate based on prompt properties alone. Further, there is no requirement from these data for any progenitor (or afterglow) models which are distinct from those developed to explain other GRBs.

*X-ray afterglow light curve*

Observations of GRB 090423 with the *Swift* X-ray telescope began 73 seconds post burst, and were obtained principally in photon counting mode. The X-ray lightcurve is shown in Figure 3 of the main article, and is derived using the techniques of[45]. Excluding the early flare it can be fit by a broken power-law with a plateau followed by a steeper decay. with indices of, $\alpha_{1,x} = -0.04^{+0.12}_{-2.46}$ and $\alpha_{2,x} = 1.40^{+0.08}_{-0.07}$, and a break time of $t_{b,x} = 5051^{+526}_{-4052}$ seconds (all errors at 90% confidence). A further description of the lightcurve, and its comparison with the optical/IR lightcurve is provided below.

*X-ray spectroscopy*

The X-ray spectrum (created using the tools of ref 46) is adequately fit by a single power-law plus absorption due to both a fixed Galactic column and host galaxy contribution. Fitting to all the data yields $\Gamma = 1.76^{+0.09}_{-0.10}$ , while ignoring the flare provides $\Gamma = 2.05^{+0.14}_{-0.09}$. Splitting the data in segments before the flare, in the plateau and in the late decay shows no strong evidence for spectral evolution, with the afterglow spectral slope consistent with Γ=2 (β= (Γ-1) ~ 1) in each of these segments. The inclusion of the flare does result in an apparent hardening of the spectral slope, but this is not unusual in flares, and hence we will exclude the flare in subsequent analysis.

The best fitting excess column density is $N_{HI} = 1.16^{+0.50}_{-0.58} \times 10^{23}$ cm$^{-2}$ at *z*=8.3, with the flare excluded, or $N_{HI} = 6.34^{+2.76}_{-2.92} \times 10^{22}$ cm$^{-2}$ including the flare. These fits are reported as excess absorption, above the Galactic value, which was fixed in our fits at $N_{HI} = 2.89 \times 10^{20}$ cm$^{-2}$ (ref 47). To investigate the nature of possible excess absorption in the X-ray afterglow of GRB 090423 we plot in Figure S02 the observed X-ray spectrum, fitted with our best fit model. In addition we also show the improvement in the residuals with the application of an additional absorber. Because the redshift of GRB 090423 is so large the soft X-rays, which are most readily attenuated, are largely redshifted out of the XRT band, and hence even the apparently small curvature would correspond to very large column densities in the rest frame of GRB 090423. Taken at face value this suggests a column ~1 × $10^{23}$ cm$^{-2}$, which would lie at the high end of observed values for *Swift* GRBs[48]. To investigate the necessity of excess absorption we investigate the goodness of fit over the photon index – $N_{HI}$ parameter space. The results are also shown in Figure S02. Ultimately, the evidence for excess absorption appears relatively weak (~3 sigma), and may well have a rather lower value. Further, it is possible that any excess absorption could be due to additional absorption in any intervening systems at lower-z along the line of sight (again such systems are not uncommon in GRB afterglows[49]). Hence we do not consider that the apparent observation of excess absorption in GRB 090423 makes it exceptional.

Finally we note that there is very little correlation observed between the H I column inferred from X-ray absorption and that inferred from Lyman-alpha[48]. The X-ray absorption therefore has little implication for the expected host absorption in the NIR.

## DATA ANALYSIS AND INTERPRETATION

### NIR Lightcurve evolution and comparison with X-rays

As with the X-ray lightcurve the IR lightcurve of GRB 090423 apparently shows some complexity, although the IR followup is relatively sparse in comparison to the X-ray. To improve the sampling we have extrapolated all of our optical/IR photometric points to a single photometric band (the J-band) using our best-fit spectral model. We then fit this combined lightcurve, and compare the results to those obtained from the X-ray.

Both X-ray and IR exhibit a plateau phase, followed by a break to a steeper decay. Ignoring the final two IR points (see below) the decay indices in both plateau ($\alpha_{1,x} = -0.043^{+0.119}_{-2.46}, \alpha_{1,IR} = -0.001^{+0.028}_{-0.040}$) and later decay phases ($\alpha_{2,x} = 1.40^{+0.08}_{-0.07}, \alpha_{2,IR} = 1.34^{+0.21}_{-0.22}$) are consistent with each other within the fitted errors. However, the break time is apparently later in the IR than the X-ray ($t_{b,x} = 5051^{+526}_{-4052}$ s, $t_{b,IR} = 23506^{+8989}_{-2485}$ s).

At times >2 days there is an apparent flattening of the IR decay, which could also be interpreted as an optical flare. We have not included these data in the fit shown in Figure 3 of the main article. However, it is relevant to consider alternative interpretations here. Including the two data points which indicate a flare (J and K band observations taken on the night of 26 April) in the fit yields a shallower later time decay index $0.96^{+0.06}_{-0.08}$, and earlier break time $t_{b,IR} = 16262^{+2546}_{-1335}$ seconds, but a poor $\chi^2$/dof (29.97/13=2.305, compared to 7.051/11=0.641 for the fit ignoring these points). Including these data by fitting two additional power-laws (one which accounts for the flattening, and another which requires a later break as implied by limits obtained roughly 7 and 15 days post burst) improves the $\chi^2$/dof to (10.81/10=)1.081 but the probability of random improvement is non-trivial (1.4% using an F-test).

*Spectral energy distribution*

The spectral energy distribution of the GRB 090423 afterglow, and its time evolution are shown in Figure S03. In the nIR it appears very blue, with a spectral slope of index $\beta$ = 0.26 ± 0.10 (based on the average spectral fit for our first three epochs of observations (UKIRT+Gemini-N, VLT+Gemini-S and GROND). Although this is extremely blue, it is not unprecedented with some previous GRBs showing similar spectral slopes (e.g. GRB 021004[ref 50], GRB 080319B[ref 51]). There is little evidence for evolution over the time period of the observations. However, the final epoch of *J* and *K* and observations do show a redder colour than previous epochs (($J$-$K)_{early}$ = 1.34 ± 0.05 and $(J$-$K)_{late}$ =1.61 ± 0.13). Given the associated errors the evidence for reddening is marginal (~2 sigma) and we do not consider it compelling. Further, as noted above the lightcurve of GRB 090423 shows evidence for a possible flare at the time of these final epoch observations, which may explain any observed colour difference.

We additionally investigated the joint X-ray – opt/IR spectral fit[52], which is also shown in Figure S03. Our early (1.5 hour) SED suggests best fit values of $\beta_x = 0.80^{+0.06}_{-0.05}$, with $\beta_{IR} = 0.30^{+0.06}_{-0.05}$ (with a fixed $\Delta\beta$ =0.5 for a cooling break), with a cooling break at $0.02^{+0.04}_{-0.01}$ keV, no host extinction is required, $A_V < 0.08$ (the overall $\chi^2$/dof for this ft is 20.79/18 = 1.16). Should the X-ray and IR lie in different cooling regimes then the post-break decay slopes should be different by $\Delta\alpha$ = 0.25, given the large errors on the measurement of $\alpha_{2,IR}$, this is consistent with the observations. The late time fit, obtained at the time of the possible IR flare shows that the X-ray to optical spectral index has increased to $\beta_x = 0.90^{+0.10}_{-0.07}$, which is also consistent with the measure $\beta_{IR}$ at the same epoch. However, this is likely to be due to the IR flare, rather than unusually rapid motion of the cooling break.

*Photometric redshift*

The strong break between the *Y* and *J* band provides compelling evidence for the high redshift origin of GRB 090423. However, to formalize the constraints that can be obtained from the photometric observations we performed a photometric redshift fit to the available photometry. We normalized the observations to a common epoch, allowing for the fading afterglow as described in the lightcurve section. We chose and epoch centred at the time of the Gemini-South, VLT and GROND observations, and fit the resulting photometry, along with filter transmission functions in the standard *hyperz* photometric redshift code[53], on the assumption that the underlying spectrum would be well described as a power-law. At this point we also corrected our photometry for the small foreground extinction of E($B$-$V$) = 0.029. Note, that at this stage we consider the differing filter response functions. In particular, the GROND $J$-band filter extends to ~1.1 microns, while the HAWK-I $J$ has little transmission below ~1.16 microns. Hence the GROND J-band photometry will encompass Lyman-alpha for any $z$>8.04

The resulting photometric redshift fit is shown in Figure S04, and is $z = 8.06^{+0.21}_{-0.28}$, which is in excellent agreement with results obtained via direct spectroscopy. There is no requirement for any host galaxy dust extinction within the fit. We further note that this approach does not include a detailed modelling of the damping wing from either the intergalactic medium (IGM) or a host Lyman-alpha absorption, both of which would tend to deplete the blue flux around the break. The impact of including such extra parameters is relatively small (typically of order 1% in the photo-z), but would favour a somewhat higher redshift, with larger associated errors. This effect is considered in greater detail in the spectral fitting described below.

*Spectroscopic constraints on the redshift of GRB 090423*

The 1D spectra shown in Figure 2 of the main article were created by taking the fluxed and calibrated 1D spectra and binning the channels together using error weighting. The bin sizes were adjusted so that the final uncertainty on the flux in each bin was approximately the same, around 6 μJy. This can clearly be seen, with much larger bins in the parts of the spectrum with greater sky noise or lower throughput. The 2D spectra shown in that Figure were also binned into larger pixels using error weighting, but in this case the pixel size was kept constant, and the final spectra Gaussian smoothed to help visualisation.

Ideally, a firm redshift could be obtained by the location of narrow absorption lines in the afterglow spectrum. However, none were obviously visible in the data, due in part to the low signal to noise ratio. To provide a more robust search for spectral features we co-added regions of the spectrum which would contain strong absorption lines, similar to those seen in the afterglow of GRB 050904[54]. Namely the lines of SiII, CIV and OI. We then stepped through the spectrum fitting all redshifts for $8.0 < z < 8.4$ to look for any dips which would represent the presence of these co-added absorption lines. However, no significant dips were found, and hence we conclude that it is not possible to derive a line redshift from the afterglow of GRB 090423.

Although a line redshift was not possible, the strong presence of a photometric break and two independent detections of a spectroscopic break via our ISAAC and SINFONI observations (Figure 2 in the main text and Figure S05) clearly demonstrate the presence of a strong spectral break in the afterglow spectrum of GRB 090423. Having ascertained the presence of the break it is important to consider the constraints that this places on the redshift. At first sight the presence of the break at ~11400 Angstroms equates directly to a redshift of about 8.4. However, the break is not necessarily so sharp, and we cannot make firm statements about its morphology directly from our low signal to noise data. In particular, it is likely that the damping wing of Lyman-alpha is impacted by contributions from both the IGM[55] and the GRB host galaxy. Since GRB hosts frequently show strong Lyman-$\alpha$ absorption, a column of $\sim 10^{18-23}$ $cm^{-2}$ could reasonably be expected[56], although the properties of any environments at $z{\sim}8$ are so unknown that such statements are inevitably speculative.

We fitted models to the Lyman-$\alpha$ damping wing based on those described in Barkana and Loeb[39]. Fitting was performed using $\chi^2$ minimization. In this we allow both the redshift and neutral hydrogen column density in the host to be free parameters. In the fit shown in the main paper we set the neutral fraction in the IGM to be 10%, although the conclusions are fairly insensitive to this assumption, and fixed the slope and normalization of the power-law to that determined by the imaging photometry. To determine a confidence interval on the redshift determination, we conservatively assumed a flat prior on $\log(N_{HI})$ between 19 and 23, and marginalised over that parameter. A flat distribution is broadly consistent with the observed distribution of $\log(N_{HI})$ values: arguably the true distribution is somewhat peaked around 21-21.5, and starting with a prior of that sort would have narrowed the error bars somewhat on the best-fit redshift. The final error bar is asymmetric, as expected since the upper limit on the redshift is rather hard, whereas very high values for $\log(N_{HI})$ give increasingly lower values for the redshift due to the large breadth of the inferred damping wing. The best constraint on the redshift of GRB 090423 available from the ISAAC data is $z = 8.20^{+0.04}_{-0.07}$. The SINFONI spectrum (Figure S05), analysed in the same way, gives $z = 8.33^{+0.06}_{-0.11}$. This result is entirely consistent and, since it is arrived at independently, gives considerable confidence in the robustness of the result.

Combining the spectroscopy and photometry in a single analysis yields $z = 8.26^{+0.07}_{-0.08}$, which we regard as our best estimate of the redshift of GRB 090423. To illustrate the insensitivity of the redshift to assumptions about the state of the IGM ionization, if we take a 100% neutral fraction we find $z = 8.22^{+0.05}_{-0.05}$.

| **Time of obs (UT)** | **ΔT(s)** | **Telescope** | **Inst** | **Filter** | **Exp (s)** | **Magnitude** | **Flux** |
|---|---|---|---|---|---|---|---|
| 23 April 07:55:59 | 40 | BOOTES-3 | - | Clear | 2 | >14.5 | <4570 |
| 23 April 07:55:59 | 300 | BOOTES-3 | - | Clear | 84 | >18.5 | <115 |
| 23 April 07:58:37 | 228 | P60 | - | r’ | 60 | >21.9 | <6.4 |
| 23 April 07:58:37 | 427 | P60 | - | r’ | 180 | >22.3 | <4.2 |
| 23 April 07:58:37 | 1268 | P60 | - | r’ | 780 | >22.6 | <3.3 |
| 23 April 08:00:02 | 313 | P60 | - | i’ | 60 | >21.4 | <9.6 |
| 23 April 08:00:02 | 930 | P60 | - | i’ | 300 | >22.1 | <5.1 |
| 23 April 08:00:02 | 1497 | P60 | - | i’ | 660 | >22.6 | <3.4 |
| 23 April 08:01:09 | 399 | P60 | - | z’ | 60 | >19.8 | <41.7 |
| 23 April 08:01:09 | 1016 | P60 | - | z’ | 300 | >20.5 | <23.3 |
| 23 April 08:01:09 | 1605 | P60 | - | z’ | 660 | >21.1 | <13.6 |
| 23 April 08:16:32 | 1492 | UKIRT | WFCAM | K | 360 | 17.96 ± 0.04 | 43.48 ± 1.47 |
| 23 April 08:23:51 | 1961 | UKIRT | WFCAM | K | 360 | 17.94 ± 0.04 | 44.45 ± 1.46 |
| 23 April 08:32:09 | 2431 | UKIRT | WFCAM | K | 360 | 18.01 ± 0.04 | 41.64 ± 1.50 |
| 23 April 09:11:06 | 4780 | Gemini-N | NIRI | Y | 480 | >22.5 | 0.27 ± 0.70 |

| **Time of obs (UT)** | **ΔT(s)** | **Telescope** | **Inst** | **Filter** | **Exp (s)** | **Magnitude** | **Flux** |
|---|---|---|---|---|---|---|---|
| 23 April 09:25:03 | 5560 | Gemini-N | NIRI | J | 480 | 19.24 ± 0.03 | 33.70 ± 1.01 |
| 23 April 09:39:27 | 6520 | Gemini-N | NIRI | H | 480 | 18.57 ± 0.02 | 36.60 ± 0.77 |
| 23 April 17:33:35 | 37790 | 6.0m BTA | SCORPIO | Ic | 3600 | >24.7 | <0.29 |
| 23 April 23:08:00 | 60120 | ESO2.2m | GROND | g | 4000 | >25.0 | <0.36 |
| 23 April 23:08:00 | 60120 | ESO2.2m | GROND | r | 4000 | >25.1 | <0.33 |
| 23 April 23:08:00 | 60120 | ESO2.2m | GROND | i | 4000 | >24.2 | <0.76 |
| 23 April 23:08:00 | 60120 | ESO2.2m | GROND | z | 4000 | >24.0 | <0.91 |
| 23 April 23:08:00 | 60120 | ESO2.2m | GROND | J | 4000 | 20.66 ± 0.08 | 9.11 ± 0.55 |
| 23 April 23:08:00 | 60120 | ESO2.2m | GROND | H | 4000 | 19.94± 0.08 | 10.40 ± 0.83 |
| 23 April 23:08:00 | 60120 | ESO2.2m | GROND | K | 4000 | 19.36 ± 0.12 | 11.20 ± 1.22 |
| 23 April 23:38:21 | 58566 | Gemini-S | GMOS | z | 3000 | >26.1 | -0.005 ± 0.043 |
| 23 April 23:28:41 | 56001 | VLT | HAWKI | K | 2640 | 19.18 ± 0.03 | 13.20 ± 0.40 |
| 24 April 00:31:36 | 59777 | VLT | HAWKI | J | 1200 | 20.52 ± 0.04 | 10.40 ± 0.42 |
| 24 April 01:00:30 | 61511 | VLT | HAWKI | Y | 2640 | >25.0 | -0.08 ± 0.09 |
| 24 April 01:38:52 | 67436 | ESO2.2m | GROND | J | 4000 | 20.76 ± 0.07 | 8.31 ± 0.58 |
| 24 April 01:38:52 | 67436 | ESO2.2m | GROND | H | 4000 | 20.00 ± 0.07 | 9.81 ± 0.69 |
| 24 April 01:38:52 | 67436 | ESO2.2m | GROND | K | 4000 | >19.8 | <7.37 |
| 24 April 05:30:34 | 79695 | UKIRT | WFCAM | K | 3600 | 19.95 ± 0.14 | 6.49 ± 0.91 |
| 24 April 23:36:36 | 142877 | VLT | ISAAC | J | 3600 | 21.80 ± 0.18 | 3.19 ± 0.51 |
| 27 April 01:04:05 | 320926 | VLT | HAWKI | J | 1200 | 22.19 ± 0.11 | 2.44 ± 0.27 |
| 27 April 00:01:14 | 316531 | VLT | HAWKI | K | 2640 | 20.58 ± 0.06 | 3.63 ± 0.22 |
| 30 April 23:06:33 | 661077 | VLT | HAWKI | J | 2400 | >23.2 | 0.44 ± 0.24 |
| 8 May 23:29:09 | 1353513 | VLT | HAWKI | K | 2640 | >21.9 | 0.22 ± 0.33 |

Table S1: Photometric observations of the GRB 090423 afterglow. The UT date refers to the start time of the observations, ΔT, is the mid-point of the observations, measured after the trigger time, which is 23 April 2009, 07:55:19.3 UT. The exposure time refers to the total on-source observation time. Magnitudes have not been corrected for the small foreground extinction of E(B-V) = 0.029.

| Telescope | Instrument | λ-range (μm) | Time (UT) | ΔT (s) | Exp(s) | Resolution |
|---|---|---|---|---|---|---|
| VLT Antu (UT1) | ISAAC(SJ) | 1.10-1.40 | 24 Apr 01:30:55 | 63335 | 2160 | 550 |
| VLT Antu | ISAAC(SZ) | 0.98-1.10 | 24 Apr | 67205 | 2700 | 500 |
| (UT1) | | | 02:35:24 | | | |
| VLT Yepun(UT4) | SINFONI(J) | 1.10-1.40 | 24 Apr 23:12:21 | 141422 | 9000 | 2000 |

Table S3: Log of spectroscopic observations of the GRB 090423 afterglow. The wavelength range of each spectrum is shown, along with its start time and resolution. ΔT refers to the time since burst at the start of the observations

| | | |
|---|---|---|
| **Fluence (15-150 keV)** | (5.9 ± 0.4) × $10^{-7}$ ergs $cm^{-2}$ | |
| **Peak Energy ($E_p$)** | 82 ±15 keV (Fermi)<br>48.6 ± 6.2 (BAT) | Von Kienlin GCN 9229 |
| **Rest frame peak energy** | 754 ± 138 keV (Fermi)<br>451 ± 58 (BAT) | |
| **Isotropic equivalent energy ($E_{ISO}$)** | (1.0 ± 0.3) × $10^{53}$ ergs (8-1000 keV) | Von Kienlin GCN 9229, 9251 |
| **X-ray afterglow photon index** | $2.05^{+0.14}_{-0.089}$ | |
| **X-ray flux at 10 hours (0.3-10 keV)**<br>**X-ray flux at 10 hours restframe** | 2.0 × $10^{-13}$ ergs $s^{-1}$ $cm^{-2}$<br>2.5 × $10^{-14}$ ergs $s^{-1}$ $cm^{-2}$ | |
| **X-ray luminosity at 10 hours (restframe)** | $2.1 \times 10^{46}$ ergs $s^{-1}$ | |

Table S3: Basic prompt and afterglow properties of GRB 090423 as derived from the *Swift* BAT and XRT data, and our later time optical/IR observations. Additional parameters from the Fermi GBM have also been reported.

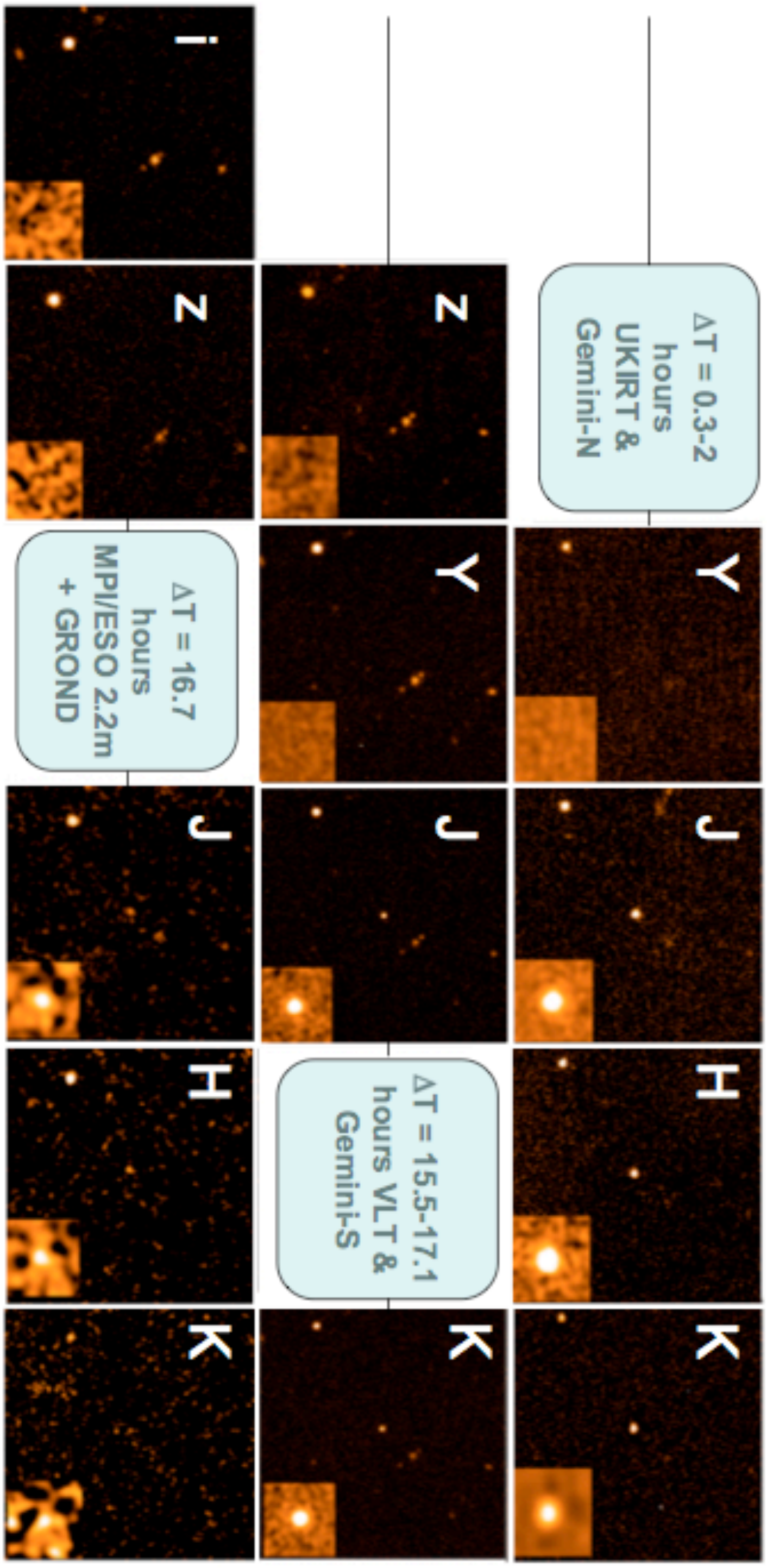

**Figure S01**: Mosaic showing multi-colour images of GRB 090423 obtained by UKIRT, Gemini-N, Gemini-S, the VLT and the MPI/ESO 2.2m over the first 24 hours post burst. All the data are consistent in showing a strong break between the Y and J-bands, and allow us to determine a photometric redshift of $z = 8.06^{+0.21}_{-0.28}$.

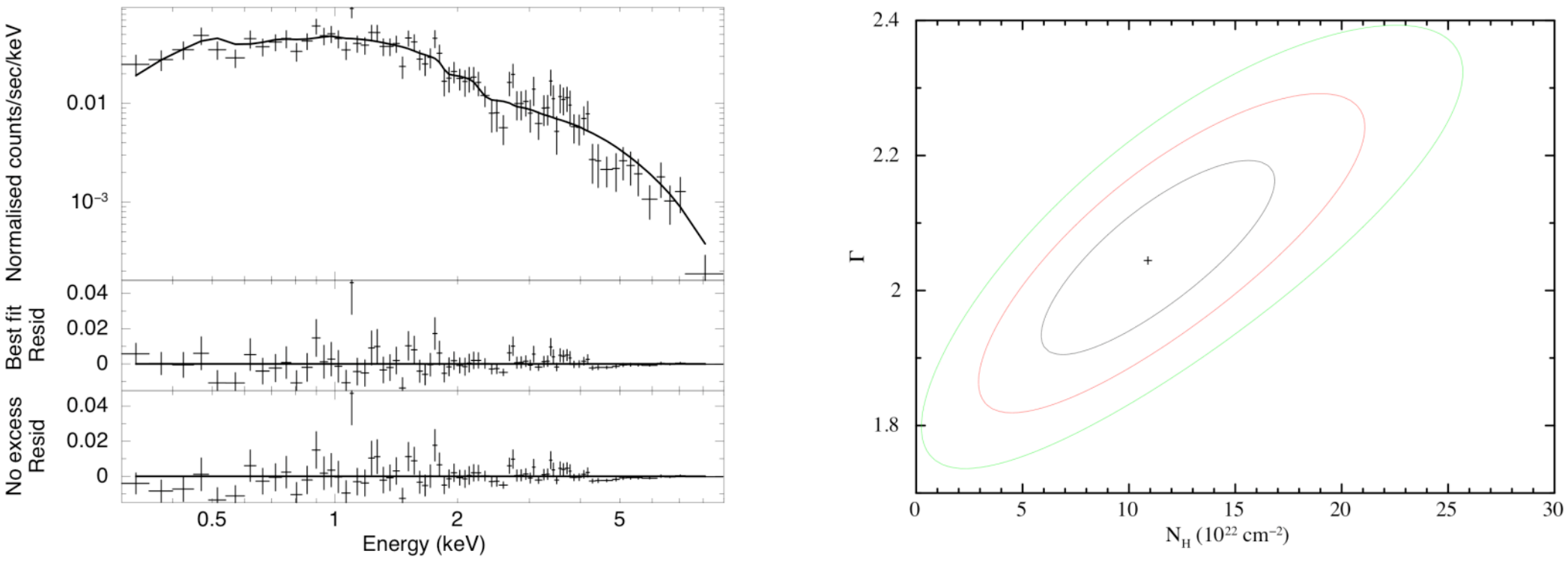

**Figure S02**: *Left:* The X-ray spectrum of GRB 090423 as observed by the *Swift* XRT. The two lower panels show the residuals to the best fit (which is shown as a solid line in the upper panel) and a fit in which zero additional extinction is included (bottom panel). This fit over-predicts the observed flux at low energies and suggests that excess column density is needed, which provides a slight improvement in the fit quality (middle panel). *Right:* Confidence contours in the $N_H$ – photon index plane for the X-ray afterglow of GRB 090423. The contours represent 1, 2 and 3 sigma confidence levels. As can be seen, allowing for a varying photon index the column is consistent with zero at ~3 sigma. This suggests that the evidence for very strong X-ray absorption is relatively weak.

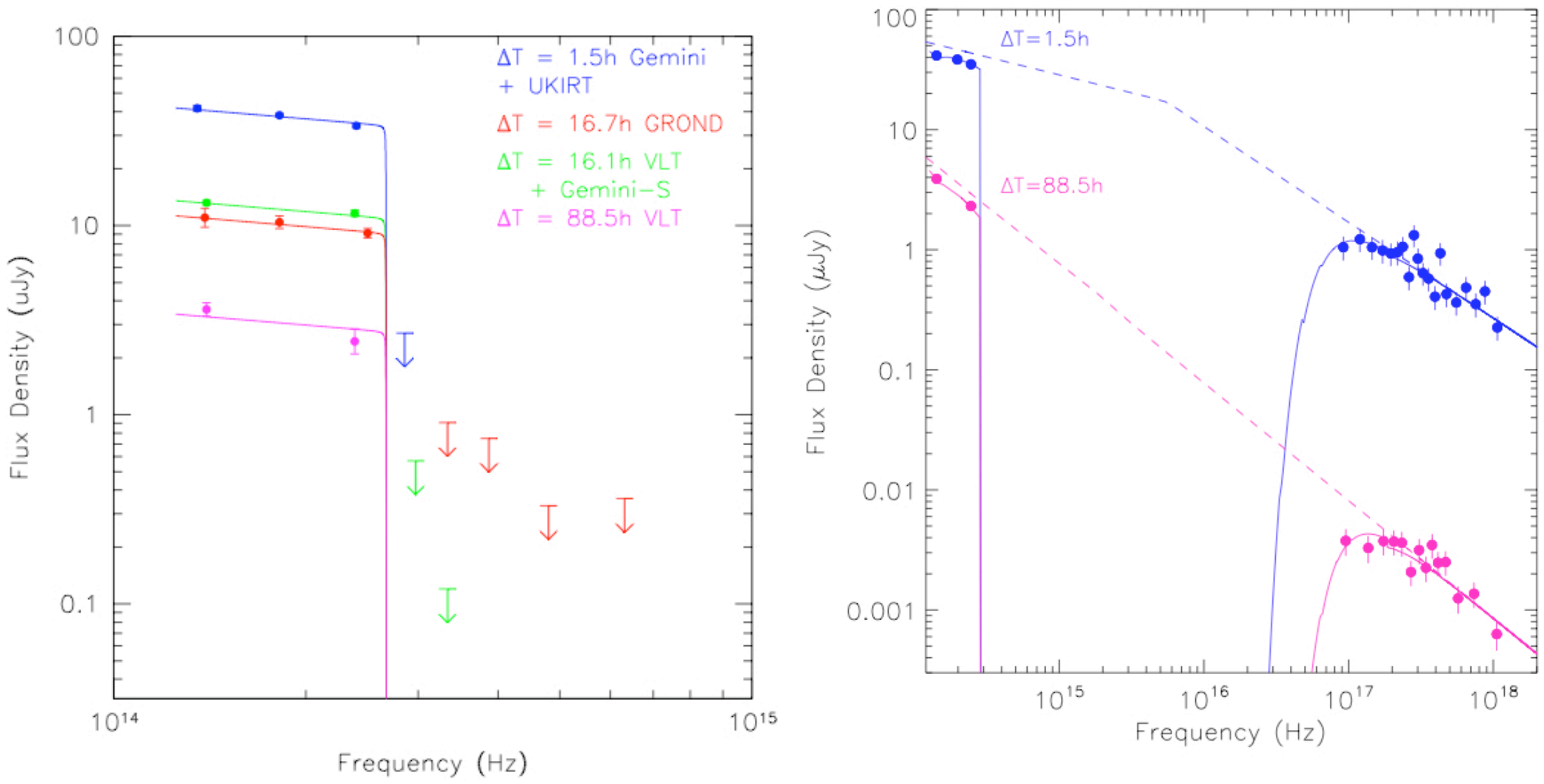

**Figure S03**: The spectral energy distribution of GRB 090423 measured at different times by our observations. The left hand panel shows only the optical/IR SED, plotted as frequency against flux density ($F_\nu$). The observations are plotted, as are the limits at the 3-sigma level, for each epoch. Our strongest constraint on the observed SED comes from our VLT observations taken ~16.1 hours after the burst. In addition we have also plotted a model for the GRB afterglow, which has β=0.27, and a redshift of $z$=8.27. The right hand panel shows the optical to X-ray SED at two snapshots, 1.5 and 90 hours post burst. This implies the cooling break lies between the X-ray and the optical. The late time SED is likely impacted by the presence of a flare in the IR rather than unexpectedly rapid motion of the cooling break.

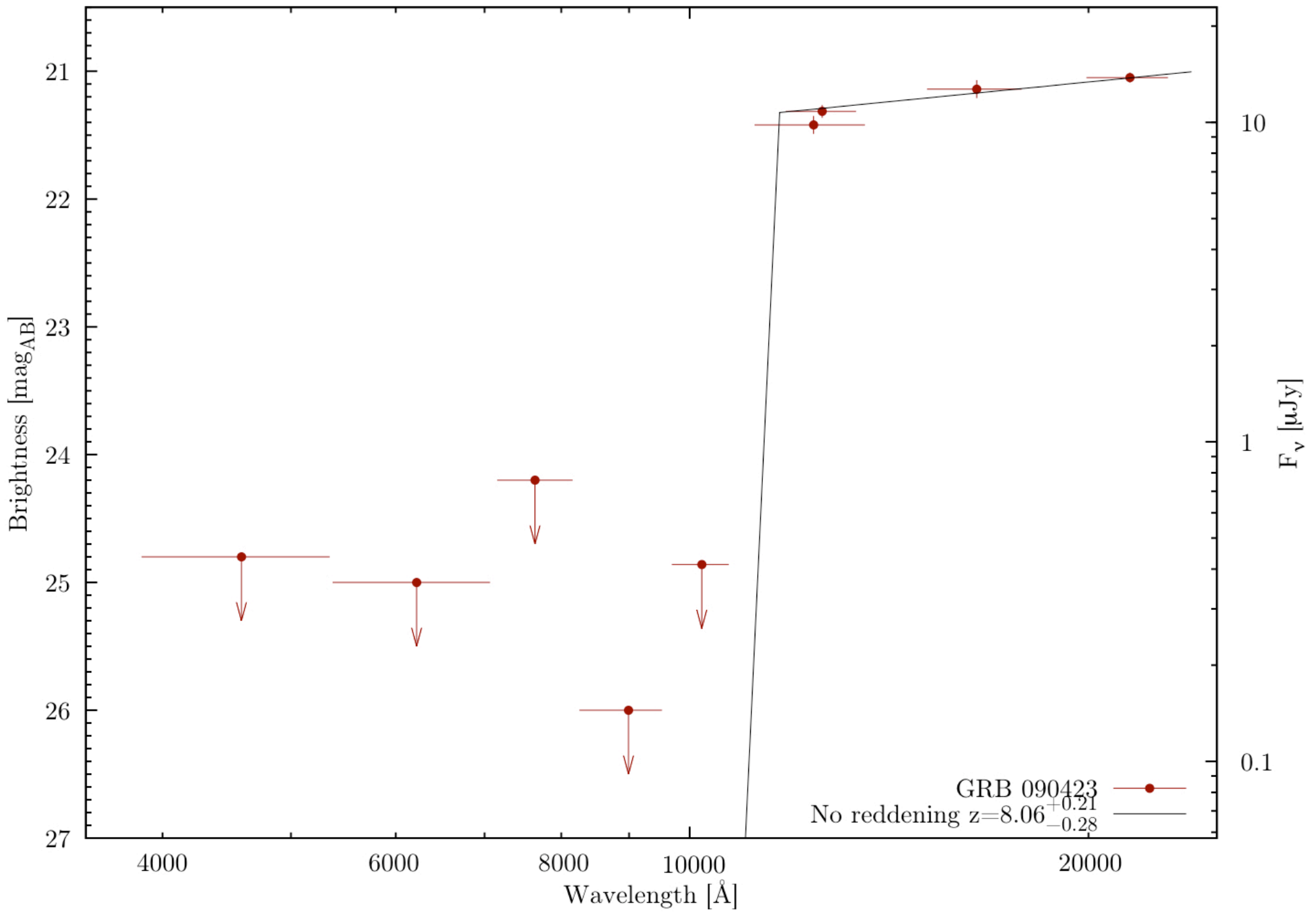

**Figure S04**: A photometric redshift fit for the afterglow of GRB 090423. The SED has been constructed at time t=16 hours, and makes use of constraints obtained from GROND, Gemini South Multi-Object Spectrograph (GMOS-S) and VLT High Acuity Wide field K-band Imager (HAWKI). The photo-z is found to be $z = 8.06^{+0.21}_{-0.28}$ in good agreement with the values obtained via spectroscopy.

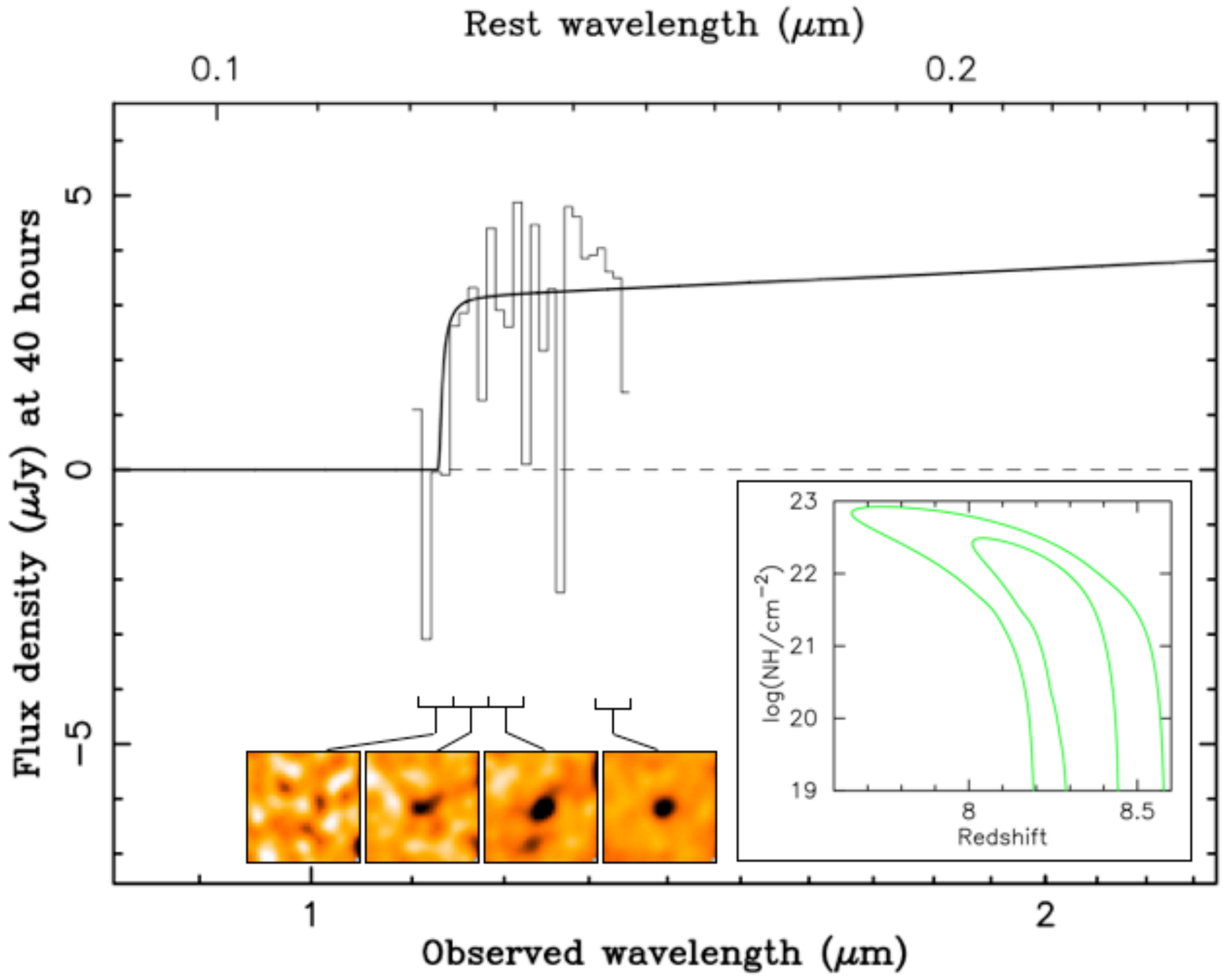

**Figure S05**: Our SINFONI IFU spectroscopy observations of GRB 090423, set on the same scale, and with the same overplotted model, as Figure 2 from the main journal text for ease of comparison. The data were reduced as described in the text. The spectral bins are 100 Angstrom wide, and the images show individual 400 Angstrom wide sections of the data-cube that have been co-added (we avoid the regions worst affected by high sky noise). The afterglow is clearly visible long-ward of ~1.14 microns, but shows no evidence of detection short-ward of this value. As for Figure 2, the 68% and 95% confidence contours are shown, and provide excellent agreement with the redshift determined from our ISAAC spectroscopy, and an independent confirmation of this value.

**SUPPLEMENTAL MATERIAL : REFERENCES**